# Crypto Technology - Impact on Global Economy

**Arunkumar Pillai***

*Central Washington University, United States*

*****Corresponding Author**
Arunkumar Pillai, Central Washington University, United States.

## 1. Introduction and Background

The last decade has been marked by the evolution of cryptocurrencies, which have captured the interest of the public through the offered opportunities and the feeling of freedom, resulting from decentralization and lack of authority to oversee how cryptocurrency transactions are conducted. The innovation in crypto space is often compared to the impact internet had on human life. There is a new term called Web 3.0 for denoting all new computing innovations arising due to the blockchain technologies. Blockchain has emerged as one of the most important inventions of the last decade with crypto currencies or financial use case as one of the domains which progressed most in the last 10 years. It is very important to research about Web 3 technologies, how it is connected to crypto economy and what to expect in this field for the next several decades.

Economics research so far has provided little insight into the economic relevance of cryptocurrencies. Most existing models of cryptocurrencies are built by computer scientists who mainly focus on the feasibility and security of these systems. Crucial issues such as the incentives of participants to cheat and the endogenous nature of some key variables such as the real value of a cryptocurrency in exchange have been largely ignored. Such considerations, however, are pivotal for understanding the optimal design and, hence, the economic value of cryptocurrency as a means of payment. Crypto currency uses an underlying Blockchain technology which solves an important double spend problem for financial industry. The original Bitcoin blockchain solved the double spend problem using a costly distributed computing transaction validation process also known as mining. Blockchain technologies have evolved to handle many complex problems and one of the most popular blockchain Ethereum can be looked at as a powerful computer that can solve several day-to-day problems. This paper analyses crypto industry's overall contribution to economy thus far and what areas it will continue to contribute and make significant economical contributions. The analysis will include the potential newer innovations that are still in early stages and role of regulators to help accelerate the industry.

## What is the Impact of Cryptocurrencies on Global Economy?

The growth of the digital assets industry, particularly in the last 12 years with the rise of Bitcoin, has been unprecedented. It has not only become larger than many countries in a short period of time, but it has also created numerous job and economic opportunities. The majority of these jobs were created in the aftermath of the industry's boom in 2017. From 2015 to 2019, the percentage of crypto-related jobs in the mainstream economy increased by nearly 1,500%. There are currently over 17,000 different cryptocurrencies being traded on 459 exchanges, with a total value of about $1.7 trillion.

Consider that the applications of blockchain technology have greatly expanded in recent years. Non-fungible tokens (NFTs) for digital art sales, decentralized autonomous organizations (DAOs) for fundraising, and play-to-earn games with blockchain features are all recent innovations. These new applications further complicate the sector and increase the demand for new talent [1].

Technology jobs in crypto industry grew 395% in the U.S. from 2020 to 2021, outpacing the wider tech industry, which saw a 98% increase [2]. There is also a surge in the influx of funding in the industry. Investors worldwide poured $30 billion into crypto and blockchain startups in 2021, according to Pitchbook data. At the same time, public interest in crypto exploded as high profile evangelists like Elon Musk praised the technology—and crypto companies entered the mainstream, as evidenced by the newly-christened Crypto.com Arena in Los Angeles.

Institutional economics understands the economy as made of rules. Rules (like laws, languages, property rights, regulations, social norms, and ideologies) allow dispersed and opportunistic people to coordinate their activity together. Rules facilitate exchange — economic exchange but also social and political exchange as well. What has come to be called crypto economics focuses on the economic principles and theory underpinning the blockchain and alternative blockchain implementations. It looks at game theory and incentive design as they relate to blockchain mechanism design. Institutional crypto economics is interested in the rules that govern ledgers, the social, political, and economic institutions that have developed to service those ledgers, and how the invention of the blockchain changes the patterns of ledgers throughout society. Institutional crypto economics gives us the tools to understand what is happening in the blockchain revolution — and what we can't predict.

Blockchains are an experimental technology. Where the blockchain can be used is an entrepreneurial question. Some ledgers will move onto the blockchain. Some entrepreneurs will try to move ledgers onto the blockchain and fail. Not everything is a blockchain use case. We probably haven't yet seen the blockchain killer app yet. Nor can we predict what the combination of ledgers, cryptography, peer to peer networking will throw up in the future. This process is going to be extremely disruptive. The global economy faces (what we expect will be) a lengthy period of uncertainty about how the facts that underpin it will be restructured, dismantled, and reorganized.

The best uses of the blockchain must be 'discovered'. Then they have to be implemented in a real world political and economic system that has deep, established institutions that already service ledgers. That second part will not be cost free.   Best Analogy would be that of how TikTok a social media application using short videos revolutionized the entire social media land scape using their inventive approach.  Blockchain technology brings a significant paradigm shift to internet. It now elevates the internet community to not only interact, share, comment on original content, but provides a fool proof model for original content ownership. This model can be a huge once an application developer solves an interesting real-life use case using this technology.

The blockchain and associated technological changes will massively disrupt current economic conditions. The industrial revolution ushered in a world where business models were predicated on hierarchy and financial capitalism. The blockchain revolution will see an economy dominated by human capitalism and greater individual autonomy. How that unfolds is unclear at present. Entrepreneurs and innovators will resolve uncertainty, as always, through a process of trial and error. No doubt great fortunes will be made and lost before we know exactly how this disruption will unfold.

## What will be the Major Innovations Using Blockchain Technology?

From financial services to retail and e-commerce, media and entertainment, healthcare, IT, government, and energy: Almost every sector is expected to adopt Web3.0 blockchain. Because Web3.0 relies heavily on blockchain, many wrongly believe that its fate is inevitably linked to the volatile cryptocurrency market. However, cryptos are just one part of the new sector [3]. Gartner explains that while cryptocurrencies crashed in the first half of 2022, decisionmakers should not assume that the value of Web3.0 technology is affected. According to the research and consulting firm, Web3.0 tech will soon reach its adoption tipping point and industries from aircraft maintenance to food safety will tokenize their applications. Web3 blockchain will completely transform the existing conventional processes of the different sectors.  In 2023, the Web3.0 blockchain technology sector will be worth more than six trillion dollars, and Web3.0 will continue to grow at a CAGR of 44.6% from 2023 to 2030 [4]. The concept of Web3.0 implies data ownership and decentralized control. The first version of the internet, Web1.0, was built solely on content produced by governments, organizations, and businesses. This web was mainly oriented to information and slowly but gradually shifted to a consumer-driven space [5].

DAOs are a type of organizational structure that have been built through the power of blockchain. DAOs are a group of like-minded individuals who work together for the long-term success of the project or creator that they are backing. DAOs offer "a say" with voting rights usually offered if you are a token holder of the project that you're a member of [6].

## 2. Growth of Crypto Economy

Research using existing data available in another research.  We use the market capitalization of crypto currencies and coins operating in the Web 3 technologies.  The following diagram will help to understand the market cap growth of Crypto economy in last decade. Market cap is a good indicator for measuring the product value of an industry.

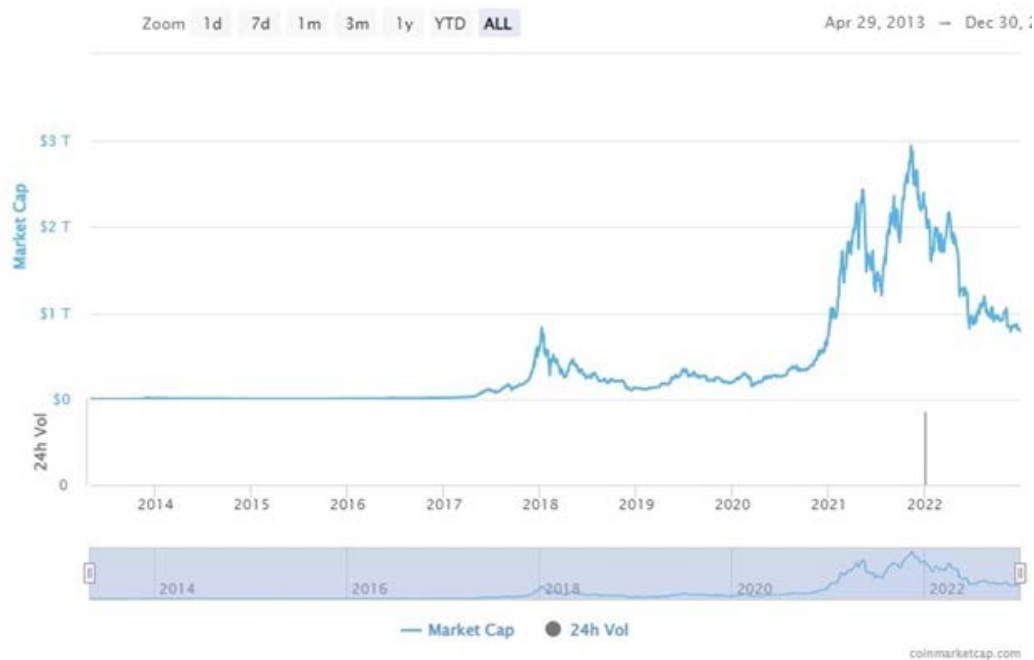

**Figure 1:** Crypto currency market cap comparison

The digital assets industry has experienced rapid growth in the past several years, reaching a market capitalization of $3 trillion in November 2021. However, the past year has been difficult for the industry, with a 70% drop in market cap to around $790 billion in December 2022. While this decline may appear concerning, it is important to consider the industry's performance within the larger context of the global economy. It is possible that the market is behaving similarly to the high-tech industry and may be preparing for the next phase of growth. Overall, the data suggests that there is still significant potential for growth in the digital assets industry.

Using the secondary data analysis, one can investigate a successful business use case solved by block chain technologies by Ripple Labs, US based block chain company. Native use case of cross border payments using Web 3 technology. Ripple which owns a coin in the symbol of XRP is a very successful crypto protocol which has established a unique business use case of cross border payments using blockchain. Their open source, permission less and decentralized technology is carbon neutral and can settle a cross border transaction in 3-5 seconds compared to the conventional Swift transaction protocol which can take up to 3-4 business days. XRP transaction fees cost $0.0002 per transaction on average.

The global cross-border payments will reach US $156 trillion in 2022. Thus, confirming it as a trillion-dollar market. According to Juniper research, solely the B2B cross-border payments will be a $35 trillion economy in 2022 [7]. Key advantages of blockchain operated cross border payments are (a) Faster settlement – A block chain enabled cross border payment takes minutes or seconds compared to up to 5 business days for the conventional cross border payments, (b) Cost effective – transaction fees is limited to the blockchain operator, (c) Enhanced security – Blockchain cross border payments leverage the same crypto technology that is used for crypto currencies. A private key or digital signature becomes invalid in case of any hacking event, (d) Improved transparency – every transaction and holdings are easy to view on an explorer. The participants can view the transactions and entries in the system is validated.

Another secondary data to look for signals about the Web 3 industry growth would be about the private equity investments in this sector. In 2020 in average $4.8 billion were invested across web 3 and crypto related startups. In 2021 this average raised up to $31.7 billion (7 times bigger), a quick reminder that 2020-2021 were the peak of the corona crisis. In the first 2 quarters of 2022 despite the global economic crisis and war in the middle of Europe, the crypto sector Fundraised additional $15 billion which is even more than in the same quarter last year.

Though all the factors that should have negatively influenced the Web 3 sector, and the mood of investors, the industry keeps showing an exceptional growth through the last 3 years. At this time a lot of people might feel themselves unready to commit anything and develop their product with such market conditions which shows the general strength and future potential of this sector. Web3 future application land scape can be understood from the following diagram [8].

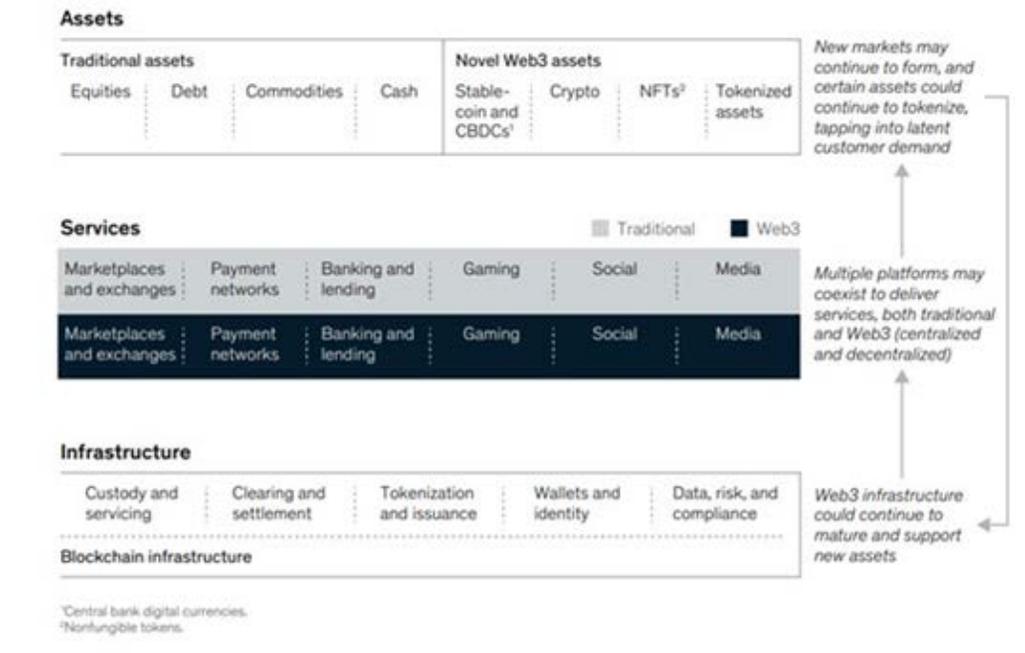

**Figure 2:** Web 3 Future

Web3 native companies are companies that utilize decentralized technologies such as blockchain and operate on the Web3 platform. These companies have the potential to disrupt a wide range of industries and create new opportunities for both themselves and incumbent companies. There are three levels at which this disruption is likely to occur: assets, infrastructure, and services. In the asset realm, new and unexplored assets such as stablecoins, non-fungible tokens (NFTs), and tokenized real estate may continue to be created and gain popularity. This presents opportunities for incumbent companies to facilitate access to these assets or tokenize their own existing assets. In terms of infrastructure, there is a need for the development of core infrastructure to support the new Web3 assets. Incumbent companies may have the opportunity to partner with Web3 native companies to innovate their offerings and support the growth of the necessary infrastructure. Finally, new Web3 native services such as marketplaces, payment networks, and deposit and loan platforms may emerge and potentially replicate the functionality of traditional services. Incumbent companies may choose to partner with these Web3 disruptors to tap into new services and bring enhanced value propositions to their user bases while retaining traditional protections and economics.

## 3. Conclusion

This paper looked at research topic of crypto technologies and how it will impact the overall global economy. It considered three potential research method to conduct the analysis and determined the secondary research method as the most effective mechanisms for conducting this research. The research looked at crypto economy market cap, the job generation in last decade in this industry and innovation landscape within Web 3 to determine that Web3 and crypto technologies are powerful disruptors in the digital industry. It has proved its value in very real business problems such as cross border payments. While it needs to be expanded to a lot more use cases, we can already see that industry is making huge progress towards the next wave of growth in decentralized finance and decentralized organizations. There is still a big gap in the regulatory frameworks that exist for this industry to strive. This paper analyzed potential measures such as registering as a state trust or national trust charter, undergoing annual audits, establishing reasonable controls and board governance, meeting basic cybersecurity standards, implementing KYC and AML policies, establishing a licensing and registration regime, requiring strong consumer protection rules, and prohibiting market manipulation [9-12].

This paper analyzed broader crypto technology industry and its impact on global economy using the past 10 years of business data. It concluded that crypto technologies have opened a new paradigm shift in digital technology using block chain powered decentralization framework. This is truly an innovation that can be compared to some of the original foundational technologies that power internet or even as basic innovation as electricity. There is a big role in this industry for improving regulatory framework which can help to make it free from fraud and bad actors trying to spoil the trust of the industry and in turn slowing the down the speed of innovation.